\documentclass[sn-basic,iicol,Numbered]{sn-jnl}% Math and Physical Sciences Reference Style
\usepackage{multirow}%
\usepackage{array}   %对表列和表格线的设置需要用到array宏包

\usepackage{graphicx}%
\usepackage{amsmath,amssymb,amsfonts}%
\usepackage{amsthm}%
\usepackage{mathrsfs}%
\usepackage[title]{appendix}%
\usepackage{xcolor}%
\usepackage{textcomp}%
\usepackage{manyfoot}%
\usepackage{booktabs}%
\usepackage{algorithm}%
\usepackage{algorithmicx}%
\usepackage{algpseudocode}%
\usepackage{listings}%
\usepackage{multirow}
\usepackage{graphicx}
\usepackage{diagbox}

\usepackage{colortbl}  %彩色表格需要加载的宏包

\usepackage{bbding}
\usepackage{multirow}

\usepackage{hyperref}
\usepackage[hyphenbreaks]{breakurl}
\usepackage{xurl}

\theoremstyle{thmstyleone}%
\theoremstyle{thmstyletwo}%

\theoremstyle{thmstylethree}%

\begin{document}

% \title[Article Title]{A General Solution to Discover More Effective Drug Candidates from the Chemical Space}

% \title[Article Title]{A General Solution to Enhance the Effectiveness of Preclinical Drug Candidate Screening

% \\

% A General Solution to Enhance the zero-shot learning
% Generalization for Drug Response Prediction}

\title[Article Title]{Zero-shot Learning of  Drug Response Prediction for Preclinical Drug Screening}


% \author*[1,2]{\fnm{First} \sur{Author}}\email{iauthor@gmail.com}



% \affil*[1]{\orgdiv{Department}, \orgname{Organization}, \orgaddress{\street{Street}, \city{City}, \postcode{100190}, \state{State}, \country{Country}}}

% \affil[2]{\orgdiv{Department}, \orgname{Organization}, \orgaddress{\street{Street}, \city{City}, \postcode{10587}, \state{State}, \country{Country}}}

% \affil[3]{\orgdiv{Department}, \orgname{Organization}, \orgaddress{\street{Street}, \city{City}, \postcode{610101}, \state{State}, \country{Country}}}

\author[]{\fnm{Kun} \sur{Li}}\email{li\_\_kun@whu.edu.cn}

\author[]{\fnm {Weiwei}
\sur{Liu}}\email{liuweiwei863@gmail.com}

\author[]{\fnm {Yong}
\sur{Luo}}\email{luoyong@whu.edu.cn}

\author[]{\fnm {Xiantao}
\sur{Cai}}\email{caixiantao@whu.edu.cn}

\author*[]{\fnm {Wenbin}
\sur{Hu}*}\email{hwb@whu.edu.cn}

\author*[]{\fnm{Bo}
\sur{Du}*} \email{gunspace@163.com}

\affil*[]{\orgdiv{School of Computer Science}, \orgname{Wuhan University}, \orgaddress{ \city{Wuhan}, \postcode{430037}, \state{Hubei}, \country{China}}}

\abstract{  Conventional deep learning methods typically employ supervised learning for drug response prediction (DRP). This entails dependence on labeled response data from drugs for model training. However, practical applications in the preclinical drug screening phase demand that DRP models predict responses for novel compounds, often with unknown drug responses. This presents a challenge, rendering supervised deep learning methods unsuitable for such scenarios.  In this paper, we propose a zero-shot learning solution for the DRP task in preclinical drug screening. Specifically, we propose a Multi-branch Multi-Source Domain Adaptation Test Enhancement Plug-in, called MSDA. MSDA can be seamlessly integrated with conventional DRP methods, learning invariant features from the prior response data of similar drugs to enhance real-time predictions of unlabeled compounds. We conducted experiments using the GDSCv2 and CellMiner datasets. The results demonstrate that MSDA efficiently predicts drug responses for novel compounds, leading to a general performance improvement of 5-10\% in the preclinical drug screening phase.  The significance of this solution resides in its potential to accelerate the drug discovery process, improve drug candidate assessment, and facilitate the success of drug discovery.
}
% enhance the effectiveness of preclinical drug candidate screening.




\maketitle

\section*{Main}\label{sec1}

\begin{figure*}[htbp]
\centering
\includegraphics[scale=0.7]{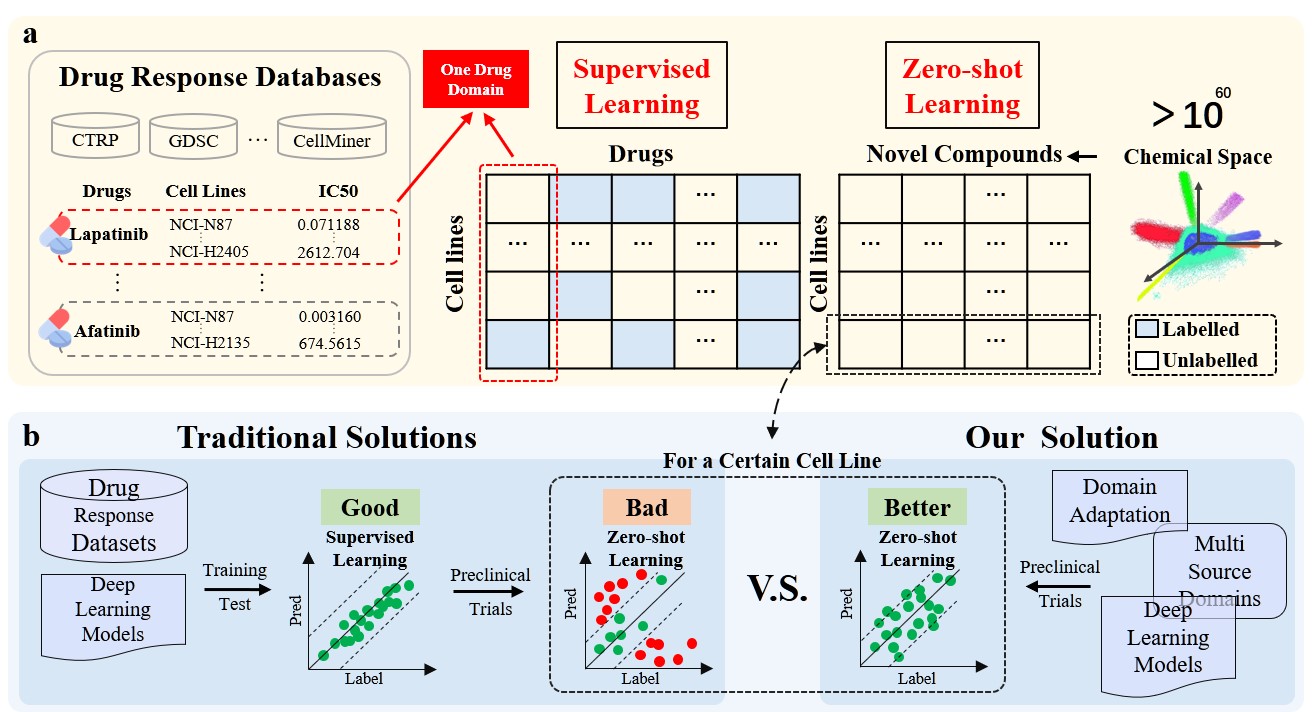}
% \caption{\textbf{The background and current status of the drug response prediction task.} \textbf{a}, Schematic on the definitions of domain, in-domain, and zero-shot learning. \textbf{b}, The comparison of the effectiveness between the existing methods and our method in preclinical experiments. The existing methods perform well in in-domain experiments but poorly in zero-shot learning environments aimed at practical applications.}
\caption{\textbf{The difference between supervised learning and zero-shot learning for preclinical drug screening.} \textbf{a}, Schematic on the definitions of supervised learning and zero-shot learning in the DRP task. \textbf{b}, The comparison of the effectiveness between the traditional solutions and our solution in preclinical trials. The traditional solutions perform well in supervised learning but poorly in zero-shot learning environments aimed at practical applications.}
\label{fig-1}
\end{figure*}

\begin{table*}[htpb]
\centering
\caption{\textbf{Examples of performance comparison for drug response prediction under the conditions of supervised learning and zero-shot learning.} Drug response data for one drug with one cell line is recorded as one sample of this drug. The sample numbers of these drugs in the training dataset are set to 0 to simulate the practical application scenario (zero-shot learning). The evaluation metric is the \textit{Pearson correlation coefficient}, and the \textbf{Global average} represents the average performance of all types of drugs in the dataset.}
\setlength{\tabcolsep}{10pt}
\renewcommand\arraystretch{1.5}
\label{tab:example}
\resizebox{\textwidth}{!}{%
\begin{tabular}{cccccccc}
\\ \hline
 \multirow{2}{*}{ \textbf{Drugs} } & \multirow{2}{*}{ \textbf{Conditions for learning} } & \multicolumn{2}{c}{\textbf{Number of samples}} & \multicolumn{2}{c}{\textbf{GraphDRP}} & \multicolumn{2}{c}{\textbf{GratransDRP}} \\ 
 &  & Train & Test & Origin & +MSDA & Origin & +MSDA \\ \hline
\multirow{2}{*}{5-Fluorouracil} & Supervised  & 90.28\% & 9.72\% & 0.9137 & \textbf{-} & 0.8878 &  \textbf{-}\\
 & Zero-shot  & \textbf{-} & 100\% & 0.465 & 0.6513 (\textcolor[RGB]{200,0,0}{ 40.1\%$ \uparrow$}) & 0.5782 & 0.6501 (\textcolor[RGB]{200,0,0}{ 12.4\%$ \uparrow$}) \\ \hline
\multirow{2}{*}{Pelitinib} & Supervised  & 88.62\% & 11.38\% & 0.8864 & \textbf{-} & 0.8884 &  \textbf{-}\\
 & Zero-shot  & \textbf{-} & 100\% & 0.3395 & 0.5887 (\textcolor[RGB]{200,0,0}{ 73.4\%$ \uparrow$}) & 0.4491 & 0.5789 (\textcolor[RGB]{200,0,0}{ 28.9\%$ \uparrow$}) \\ \hline
\multirow{2}{*}{Alectinib} & Supervised  & 89.58\% & 10.42\% & 0.7015 & \textbf{-} & 0.8568 &  \textbf{-}\\ 
 & Zero-shot  & \textbf{-} & 100\% & 0.1424 & 0.4224( \textcolor[RGB]{200,0,0}{ 196.6\%$ \uparrow$}) & 0.2581 & 0.4149 (\textcolor[RGB]{200,0,0}{ 60.8\%$ \uparrow$}) \\ \hline
\multirow{2}{*}{Global average} & Supervised  & 88.89\% & 11.11\% & 0.9271 & \textbf{-} & 0.9342 & \textbf{-} \\
 & Zero-shot  & \textbf{-} & 100\% & 0.4402 & 0.4898 (\textcolor[RGB]{200,0,0}{ 11.3\%$ \uparrow$}) & 0.4848 & 0.5103(\textcolor[RGB]{200,0,0} {5.3\%$ \uparrow$}) \\ \hline
\end{tabular}%
}
\end{table*}

Improving the efficiency of preclinical drug candidate screening is a long-standing core challenge in the field of drug discovery \cite{NMI_4}. It takes an average of 10 to 15 years and more than \$2 billion for a new drug to reach the pharmacy shelf \cite{OverviewOfDD}. Historically, natural products have been the main source of new drug entities; in recent years, however, there has been a shift towards high-throughput screening (HTS) techniques \cite{Nature_1, HTS_1}. HTS drug screening methods can screen chemical libraries to identify the most promising compounds that have the desired effect on specific biological targets. Notably, the conventional libraries employed in HTS and virtual ligand screening  \cite{ VLS_3} (VLS) are constrained to less than 10 million accessible compounds, representing a mere fraction of the vast existing chemical space encompassing an estimated $10^{20}$ to $10^{60}$ novel compounds \cite{howmanydrugs}. This limitation of standard HTS and VLS decelerates the pace of drug discovery \cite{lyu2019ultra, stein2020virtual}, frequently yielding compounds with moderate affinity, limited selectivity, and initial hits displaying absorption, distribution, metabolism, excretion, and toxicity (ADMET) factors profiles \cite{ADMET} that necessitate extensive testing.

In recent years, driven by the rapid advancement of AI technology, virtual screening based on deep learning (DL) is poised to emerge as a swifter and more cost-effective approach to drug discovery \cite{DRforCOVID19, NMI_2}. Exhaustive catalogs of somatic mutations in various cancer types have been created \cite{firstG1, firstG2} and major oncogenic mutations identified \cite{firstG3}. As a result, the establishment of cancer drug sensitivity databases, such as the Genomics of Drug Sensitivity in Cancer (GDSC) \cite{GDSC},  has made a large amount of drug response data newly available to researchers. Many approaches leverage these databases to design and validate a diverse array of models that aim to elucidate connections between genomic data and drug responsiveness in the context of drug discovery \cite{NMI_3}.

% 近年来，随着AI技术的快速发展，依靠深度学习进行虚拟筛选有望成为更快、更便宜的药物发现方法。各种癌症类型的体细胞突变的详尽目录已经创建\cite{firstG1，firstG2}，并且主要的致癌突变已被识别\cite{firstG3}。 因此，针对癌症药物的敏感性数据库的创建，例如癌症药物敏感性基因组学（GDSC）\cite{GDSC}，提供了大量真实的药物反应数据。 许多方法利用这些数据库设计并验证了各种各样的模型提取基因组信息和药物反应性之间的联系的能力，以进行药物发现。

% 先前的研究在域内实验的方法设计取得了良好的结果，并表明了药物再利用方面具有潜在的应用\cite{chang2018cancer, mainDRP1, mainDRP2}。 然而，这些模型的实际应用表明，当面对未知的药物数据域（即域外）时，预测准确性会降低\cite{GraphDRP，tcnns}。

% 这意味着先前的方法无法准确预测未在训练集中出现的无标签的类药化合物 \cite{DRPTalkODD1, mainDRP4, smalley2017ai}。 如何通过有限的已知药物数据域准确高效地预测无标签的类药化合物的药物反应?如何提高DRP模型的泛化性?这个问题已成为药物发现在现有DRP任务的重点。

% The distribution of data between the response data of known drugs and the response data of novel compounds is inconsistent, which makes supervised learning DRP methods ineffective in predicting novel compounds, as shown in Fig. \ref{fig-1} b.

Conventional drug response prediction (DRP) methods typically rely on supervised learning using labeled response data from the drugs \cite{chang2018cancer, mainDRP1, mainDRP2}, as depicted in Fig. \ref{fig-1}(a). Predicting the responses of unlabeled novel compounds constitutes a zero-shot learning challenge \cite{TPAMI_Zero_Shot_Learning, TPAMI_Zero_Shot_Learning2} 
since these novel compounds may not have been encountered during training. Furthermore, the drug responses of novel compounds in practical preclinical drug screening applications are unknown. The distribution of data between known drug response data and that of novel compounds is often inconsistent, rendering supervised learning-based drug response prediction methods ineffective for novel compounds  \cite{DRPTalkODD1, mainDRP4, smalley2017ai}, as illustrated in Fig. \ref{fig-1}(b).  To underscore the shortcomings of supervised learning methods in practical preclinical drug screening, we test two state-of-the-art (SOTA) DRP methods (GraphDRP \cite{GraphDRP}, GratransDRP \cite{GraTransDRP}) on three drugs  (5-Fluorouracil, Pelitinib, Alectinib), as shown in Table \ref{tab:example}. These methods typically achieve correlation metrics exceeding 90\% in a supervised learning experimental setup but exhibit a significant drop to only 20-40\% correlation in a zero-shot learning scenario.  The experimental results indicate that these methods are unable to accurately predict novel compounds in the practical application of preclinical drug screening. How, then, can the drug responses of previously unseen novel compounds be effectively predicted? This zero-shot learning problem has become the focus of DRP tasks in preclinical drug screening. 

 % that do not appear in the training dataset

There are many research areas related to zero-shot learning, such as domain generalization, meta-learning, transfer learning, covariate shifting, and so on. Recently, \textit{Domain Adaptation} (DA) has gained popularity in the machine learning field due to its demonstrated effectiveness in enhancing model prediction performance, especially when dealing with unlabeled test samples.  This applicability extends to domains like computer vision \cite{CVPR20_DA_ForSemantic_Segmentation, CVPR20_DA_ForSemantic_object_detection, CVPR20_DA_ForSemantic_Action_Recognition}, as well as  natural language processing \cite{NLPDA1, NLPDA2, NLPDA3}. As shown in Fig. \ref{fig-1}(a), each drug can interact with numerous known cell lines, yielding corresponding response data. We define the collection of response data for a single drug across multiple cell lines as a single drug domain.  The response data for one drug interacting with one cell line is recorded as one sample for that drug. DA aims to maximize the performance of an unknown drug domain (the \textbf{target domain}) by leveraging knowledge from known drug domains (\textbf{source domains}). This approach aligns well with the requirements of zero-shot learning in the context of the DRP task. This is due to the fact that DA is an adaptive approach, proposed with the assumption that it allows the model access to the samples in the target domain  \cite{NMI_1}. This aligns with the conditions of zero-shot learning during the testing phase of the DRP task, wherein the DRP model must handle novel compounds without access to any prior samples. Nevertheless, there are still challenges associated with effectively enhancing the zero-shot learning capabilities of DRP models for preclinical drug screening through the utilization of DA methods. These challenges include:

% The types of drugs and cell lines present in the test dataset are involved in the training dataset, such an experimental setting is defined as in-domain / supervised learning, as shown in Table \ref{fig-1}a.  In practical applications of drug discovery, the existing DRP models need to predict the response results of novel compounds (i.e., zero-shot learning / zero-shot learning). Several DA-based DRP methods \cite{NMIDRP, NMIDRP2} are proposed to predict the efficacy of the marketed anti-cancer drugs across individual patient genes  \cite{NM_PDX, DA_PDX1, DA_PDX2, DA_PDX3}, which are mainly for drug repurposing and precision medicine. The DRP task in drug discovery aims to find an effective drug structure for the treatment of a certain disease.

% Several DRP methods based on DA \cite{NMIDRP, NMIDRP2}. These methods are aimed at patient-derived xenografts (PDX) \cite{NM_PDX, DA_PDX1, DA_PDX2}, that is, predicting the efficacy of a single drug in different patients' individual genes, which is a classification task \cite{DA_PDX3}. The DRP task in drug discovery aims to predict the response of multiple drugs in a single cell line, which involves screening drugs and ranking them using a regression approach. Consequently, the DRP task in drug discovery faces several significant challenges, as outlined below:

\begin{enumerate}

% Domain adaptation (DA) \cite{DA_first} is a subfield in machine learning that aims to tackle these types of problems by adjusting for differences between domains so that a trained model can generalize to the domain of interest, which has proven effectiveness in improving model generalization on many tasks, such as computer vision \cite{CVPR20_DA_ForSemantic_Segmentation, CVPR20_DA_ForSemantic_object_detection, CVPR20_DA_ForSemantic_Action_Recognition} and natural language processing \cite{NLPDA1, NLPDA2, NLPDA3}.  In the DRP task, some scholars pay attention to the domain shift problem and propose several DRP methods based on DA \cite{NMIDRP, NMIDRP2}. These methods are aimed at patient-derived xenografts (PDX) \cite{NM_PDX, DA_PDX1, DA_PDX2}, that is, predicting the efficacy of a single drug in different patients' individual genes, which is a classification task \cite{DA_PDX3}. The DRP task in drug discovery aims to predict the response of multiple drugs in a single cell line, which involves screening drugs and ranking them using a regression approach. Consequently, the DRP task in drug discovery faces several significant challenges, as outlined below:

\item \textit{Insufficient theoretical research on domain adaptation methods for regression tasks.} Numerous effective DA methods have been devised for classification tasks \cite{TPAMI_DA1, TPAMI_DA2, DA_PDX1, DA_PDX3}. To the best of our knowledge, however, there is a paucity of methods specifically tailored to regression tasks. One perspective \cite{Jiang_2021_CVPR} posits that domain alignment may widen the margins between classes in the target domain, thereby enhancing model generalization. However, it is essential to note that the regression space is typically continuous; this can be contrasted with the classification space, in which clear decision boundaries exist. The other perspective \cite{Chen_2021_ICML} asserts that classification is robust to feature scaling but regression is not, and that aligning the distributions of deep representations would alter the feature scale and impede domain adaptation regression. To summarize, it can be concluded that domain adaptation methods pose greater challenges when applied to regression tasks compared to classification tasks.

 % in the field of drug response.

% Different from the conventional DA methods, they can roughly divide the source domain and the target domain into two or three domains; in the drug response prediction stage, the activity cliff phenomenon of the drug leads to inconsistent distribution of the response results between a drug and multiple cell lines, and the amount of drug types is huge, so the drug response data of each drug needs to be regarded as an independent drug domain, so the number of drug domains is also unprecedentedly large.

\item \textit{Numerous open-source domains, not limited source domains.}  Considering a single drug domain as the source domain for data mining results in a significant decrease in model generalization accuracy due to data availability limitations and the demands of practical application. Each drug's response data should be treated as an independent drug domain, which results in an exceptionally large number of drug domains. Furthermore, the conventional single-domain adaptation method will fail when confronted with substantial distribution differences between the source and target domains \cite{ML_DA_2010}. Therefore, the efficient construction of source domains from a pool of over 20,000 known drug domains presents a challenge due to the presence of inter-domain shifts in these source domains \cite{MDA_ECCV_1}.

\item \textit{Complex distribution patterns exist between the source and target domains.} Conventional DA methods require the alignment of only one type of input  (e.g., feature maps of images \cite{TPAMI_CV_DA1, TOG_CV_DA1}, sentence embeddings \cite{NLP_DA_NIPS, NLP_DA}).  However, the input data for the DRP task comprises two distinct types: drugs and cell lines. The types of cell lines in the source domains may not perfectly correspond to those in the target domain \cite{NMIDRP2}. Additionally, the distributions of combined features from different drugs and cell lines are inconsistent. Hence, the effective alignment of complex source and target domains is a challenging task.
 
\end{enumerate}

\begin{figure*}[h!]
\centering
\includegraphics[scale=0.50]{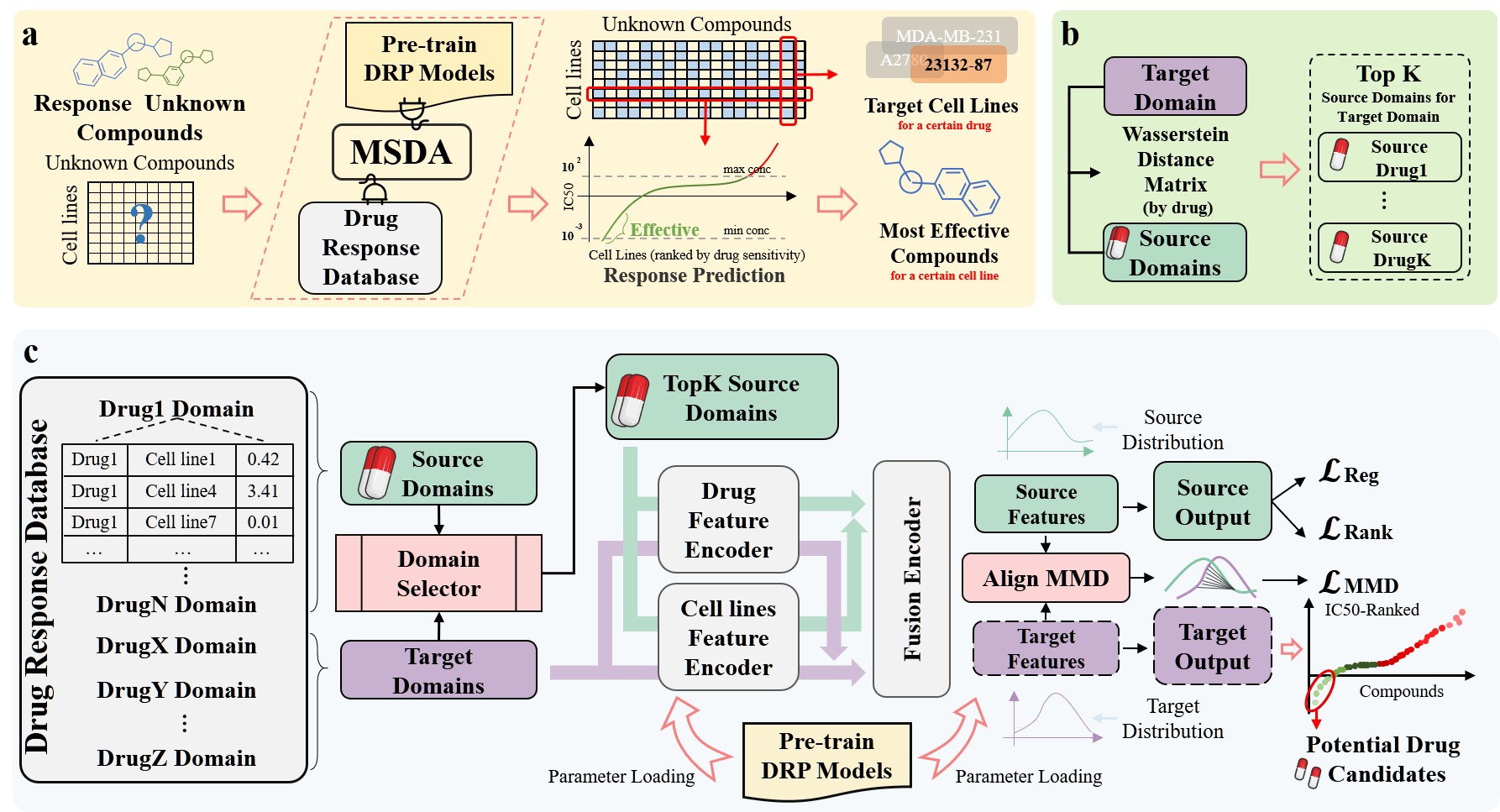}
\caption{\textbf{Overview of the our solution}. \textbf{a}, The main application of the MSDA as a plug-in for DRP tasks. \textbf{b}, A flow illustration of the domain selector, where the input is a target domain and the output is the top $K$ drug domains from the source domain with the highest similarity to the target domain. \textbf{c}, The framework of the MSDA. The input is the source domains and target domains, the output is the drug response prediction of the target domains. }
\label{fig-2}
\end{figure*}

 % In this paper, we propose a zero-shot learning solution for the DRP task in preclinical drug screening. In this solution, we propose a Multi-branch Multi-Source Domain Adaptation Test Enhancement plug-in, called MSDA.

To address the above challenges, we propose a zero-shot learning solution for the DRP task in preclinical drug screening. Within this approach, we present the Multi-branch Multi-Source Domain Adaptation Test Enhancement Plug-in (MSDA), designed to enhance the effectiveness of DRP model predictions for novel compounds. During the testing phase, the MSDA identifies source domains with the strongest correlation to the target domains. It guides the pre-trained model to acquire invariant features from these domains, facilitating adaptation to the target domain. The MSDA consists of two modules. The first module, a multi-source domain selector, employs the Wasserstein distance metric \cite{Wasserstein} on drug features to identify the most relevant drug domains from a large training dataset, treating them as multi-source domains. The second module is a multi-branch multi-source domain adaptation module based on Maximum Mean Discrepancy (MMD) \cite{MDD}. This module comprises two distinct prediction branches: the first is the prediction branch of the pre-training model itself, and the second is the target domain adaptation branch responsible for transferring multi-source domains to the target domain. The latter aligns cell line types from various domains before computing the MMD.

% To the best of our knowledge,  the MSDA is the first multi-source domain adaptation method for the generalization of DRP models.

% we propose a Multi-branch Multi-Source Domain Adaptation Test Enhancement plug-in, called msda. MSDA can be seamlessly loaded on generic drp methods to correct the prediction results of unlabelled novel compounds in real-time using existing drug response databases. 

% As a plug-and-play method, MSDA can be widely used with different DRP methods, and global experiments are carried out

% To show the extensive performance lift achieved by the MSDA, we performed exhaustive comparative studies on the MSDA variants with other state-of-the-art methods on the GDCSv2 and CellMiner. 

To showcase the substantial performance improvements attained with the MSDA, we conducted comparative studies with other SOTA methods on the GDCSv2 and CellMiner datasets. In the inference phase, the MSDA provides average improvements of 26.1\%, 15.8\%, 16.7\%, 9.3\%, 9.8\% on GDSCv2 and 26.2\%, 15.6\%, 1.5\%, 10.2\%, 11.8\% on CellMiner across four metrics for each of the five methods, namely tCNNs, DeepCDR, GraphDRP, GratransDRP, and TransEDRP, respectively. We further perform ablation studies on various aspects of the MSDA, including the strategy for the selection of source domains and the design of the target domain adaptation branch. These results highlight that MSDA achieves advanced and stable performance when dealing with unlabeled data of novel compounds. Additionally, to showcase the MSDA's potential in drug discovery, we conducted a series of hyperparameter experiments to investigate the influence of expanding the number of domain adaptation branches on performance. These experiments affirm that the MSDA plug-in in our solution is plug-and-play, highly versatile, significantly enhances generalization, and can potentially facilitate higher performance when given adequate computational resources.  The significance of this zero-shot learning solution resides in its capacity to accelerate the drug discovery process, improve drug candidate evaluation, and facilitate successful drug discovery.

\section*{Results}\label{sec2}

\subsection*{Overview of the MSDA}
\label{sec2.1}
The Multi-Source Domain Adaptation (MSDA) method, as proposed in the context of zero-shot learning for drug response prediction, seeks to identify the most efficacious potential compounds for specific cell lines (see Fig. \ref{fig-2}(a)). In practical preclinical drug screening applications, the input comprises a set of novel compounds. The MSDA is employed to forecast drug response outcomes for the drug domain associated with each compound and dynamically enhance prediction accuracy through test-time domain adaptation strategies. In the initial stage of the MSDA, a multi-source domain selector is designed to select several source drug domains similar to one target domain, as illustrated in Fig. \ref{fig-2}(b). The Wasserstein distances between the drug features of the target domain and those of the source domains are computed, after which the source domains that exhibit the highest similarity are chosen as the ultimate source domains.

The framework of the MSDA, depicted in Fig. \ref{fig-2}(c), consists of two consecutive modules. The first module is the multi-source domain selector, responsible for identifying the most relevant drug domains from the training set as multi-source domains. The second module, a multi-source domain adaptation component, comprises a drug feature representation branch and a cell line gene feature representation branch. Additionally, it includes multiple independent prediction branches with identical network structures, each of which is loaded with pre-trained model structures and weights. These prediction branches encompass the source domain prediction branch from the pre-trained model itself, as well as a minimum of one target domain adaptation branch. Each target domain adaptation branch must align with the source domain branch based on cell line types.

\begin{table*}[htpb]

\setlength{\tabcolsep}{10pt}
\renewcommand\arraystretch{1.5}
\caption{\textbf{Overall experiment.} The table shows the model performance of five representative DRP methods on both the GDSCv2 and CellMiner datasets before and after test-time domain adaptation with the MSDA. The RMSE and Rank metrics are better when they are closer to 0, while the PCC and SPC metrics, which indicate the correlation between the predicted results and the true results, are better when they are closer to 1.}
\resizebox{\textwidth}{!}{%
\begin{tabular}{cccccccccc}

\hline
  &  & \multicolumn{4}{c}{\textbf{GDSCv2}} & \multicolumn{4}{c}{\textbf{CellMiner}} \\ \hline
\multicolumn{2}{c}{Method} & PCC & RMSE &  SPC & Rank& PCC & RMSE & SPC & Rank \\ \hline
\multirow{3}{*}{tCNNs} & Original & 0.2073 & 0.0421 & 0.2032 & \textbf{0.0023} & \textbf{0.3320} & 0.00331 & \textbf{0.3627} & 0.0212 \\
  & +MSDA & \textbf{0.3709} & \textbf{0.0080} & \textbf{0.3648} & 0.0054 & 0.3309 & \textbf{0.00146} & 0.3604 & \textbf{0.0106} \\
  & (Improv.) & 79.0\% & 80.9\% & 79.53\% & -134.78\% & -0.3\% & 55.9\% & -0.63\%	& 50.00\%
 \\\hline
\multirow{3}{*}{DeepCDR} & Original & 0.4442 & 0.0048 & 0.4391 & 0.0358 & 0.3593 & 0.00028 & 0.3766 & 0.0064 \\
 & +MSDA & \textbf{0.4946  } & \textbf{0.0039} & \textbf{0.4954} & \textbf{0.0281}  & \textbf{0.3752} & \textbf{0.00024} & \textbf{0.3936} & \textbf{0.0045} \\
 & (Improv.) & 11.3\% & 17.9\% &  12.82\% & 21.51\%
  & 4.4\% & 14.3\% & 14.30\% & 29.69\%
  \\\hline
 
\multirow{3}{*}{GraphDRP} & Original & 0.4402 & 0.0056 & 0.4447 & 0.0337 & 0.4393 & 0.00027 & 0.4616 & 0.0054 \\
 & +MSDA & \textbf{0.4898} & \textbf{0.0042} & \textbf{0.4888} & \textbf{0.0263} & \textbf{0.4398} & \textbf{0.00027} & \textbf{0.4624} & \textbf{0.0051} \\
 & (Improv.) & 11.3\% & 24.0\% &  9.92\% & 	21.96\%
  & 0.1\% & 0.2\% &  0.20\% & 	5.56\%
  \\\hline
 
\multirow{3}{*}{GratransDRP} & Original & 0.4848 & 0.0050 & 0.4851 & 0.0241 & 0.4324 & 0.00026 & 0.4562 & 0.0058 \\
 & +MSDA & \textbf{0.5103} & \textbf{0.0039} & \textbf{0.5073} & \textbf{0.0224} & \textbf{0.4380} & \textbf{0.00025} & \textbf{0.4611} & \textbf{0.0042} \\
 & (Improv.) & 5.3\% & 20.6\% &  4.58\% & 	7.05\%
  & 1.3\% & 6.1\% &  6.10\% & 	27.59\%
  \\\hline
 
\multirow{3}{*}{TransEDRP} & Original & 0.5060 & 0.0052 & 0.5039 & 0.0295 & 0.4380 & 0.00025 & 0.4578 & 0.0055 \\
\multicolumn{1}{l}{} & +MSDA & \textbf{0.5316} & \textbf{0.0044} & \textbf{0.5300} & \textbf{0.0254} & \textbf{0.4422} & \textbf{0.00021} & \textbf{0.4608} & \textbf{0.0038} \\
\multicolumn{1}{l}{} & (Improv.) & 5.1\% & 15.2\% &  5.2\% & 13.9\%
  & 1.0\% & 14.7\% &  0.7\%&	30.8\%
 \\ 
\hline
\end{tabular}%
}
\label{exp:overall}

\end{table*}

\subsection*{Evaluation strategies and metrics}

We evaluate the performance of the MSDA plug-in on two publicly available datasets: GDSCv2 \cite{GDSC} and CellMiner \cite{NCI60}. The drug domain is then randomly partitioned into source and target domains in an 8:2 ratio. The DRP methods without the MSDA are trained and validated on the source domains and tested on the target domains to assess the impact of the plug-in. The MSDA focuses on the zero-shot learning problem of the DRP task. In the context of zero-shot learning experiments, it is important to note that the test dataset comprises drugs not included in the training dataset, which adds both practicality and complexity to the task. Therefore, our validation strategy is designed to simulate the practical application scenario of preclinical drug screening.  The response data is clustered by drug type and then randomly partitioned into source and target domains, with the type of drug serving as the criterion for division. It should be noted here that a single drug domain represents the smallest unit of division, as illustrated in Fig. \ref{fig-2}(c). This zero-shot learning task presents a greater level of complexity compared to the random splitting of the entire dataset commonly employed in supervised learning \cite{NMIDTI}. To comprehensively evaluate the impact of the MSDA, we employed several evaluation metrics: Root Mean Square Error (RMSE) to gauge deviation, Pearson correlation coefficient (PCC) for assessing linear correlation, Spearman's Rank Correlation Coefficient (SRC) to measure monotonicity, and Margin Ranking Loss (Rank) to evaluate ranking performance.

% we not only select the traditional model evaluation metrics Pearson's correlation coefficient (PCC) and root mean square error (RMSE) but also innovatively propose the metrics of \textit{Hit Coverage} (HC), which is a measure of the model's ability to provide effective novel compounds for the given cell lines.

% \begin{figure*}[htpb]
%     \centering
%     \includegraphics[width=0.95\linewidth]{fig4.jpg}
%     \caption{\textbf{Ablation study.} \textbf{a} and \textbf{b} are ablation experiments on target domain adaptation multi-branches and single-branches in the MSDA. In \textbf{a}, the number of multi-branches $n$ is 3 and each branch responds to 5 drug domains. In \textbf{b}, $n$ is 2 and the number of source drug domains is kept at 5. \textbf{c}, The ablation experiments on whether the target domain adaptation branches need to fuse the prediction branch of the original methods. }
%     \label{fig:Ablation}
% \end{figure*}

\subsection*{Overall experiment}

% The MSDA focuses on the problem of generalization in zero-shot learning. In the zero-shot learning experiment, drugs in the test dataset are not present in the training dataset, which is much more practical and challenging. 

Our proposed zero-shot learning solution aims to mitigate data limitations in drug response prediction,  thereby enhancing the efficiency and accuracy of drug discovery and evaluation. Our goal is to validate the effectiveness of MSDA as a test-time enhancement plug-in for implementing this solution using various datasets and diverse DRP methods. In the overall experiment, we selected five publicly available DRP methods and utilized two drug response databases. The prediction results, including those that both do and do not integrate our proposed MSDA plug-in, are presented in Table \ref{exp:overall}. The prediction results of the original models are denoted by \textbf{Original}, while results after integrating MSDA are labeled \textbf{+MSDA}. The comparative analysis demonstrates that MSDA significantly enhances the performance of DRP methods in predicting responses for unknown drug domains. Across the drug response databases GDSCv2 and CellMiner, the average PCC metrics exhibit improvements of 22.4\% and 1.3\%, respectively, while the average RMSE metrics indicate enhancements of 31.7\% and 18.2\%, respectively. This indicates that MSDA can bolster the assessment of preclinical drug candidates by offering more dependable predictions, a crucial factor for progressing potential drugs through the development pipeline. It is noteworthy that TransEDRP attains SOTA performance on both datasets, both for the original prediction results and those following MSDA integration. The MSDA and the original DRP method are decoupled, indicating that coupling MSDA with SOTA DRP methods like TransEDRP may hold even greater promise in practical applications.

\begin{table*}[htpb]
\centering
\caption{\textbf{Ablation study of the MSDA on GDSCv2 on five representative DRP methods}. \textbf{Base} indicates zero-shot learning performance of the DRP methods.  \textbf{(A)} is used to compare the effect of merging the raw prediction branch from the original methods. \textbf{(B)} and \textbf{(C)} are used to compare the effect of K and n on the performance of the MSDA.}
\setlength{\tabcolsep}{10pt}
\renewcommand\arraystretch{1.5}
\resizebox{\textwidth}{!}{%
\begin{tabular}{ccccccccc}
\hline
 &
  \multirow{2}{*}{K} &
  \multirow{2}{*}{n} &
  \multirow{2}{*}{\begin{tabular}[c]{@{}c@{}}Fusion\\ Raw Branch\end{tabular}} &
  \multicolumn{5}{c}{Result on GDSCv2   (RMSE/PCC)} \\ 
                     &                    &                    &   & tCNNs        & DeepCDR      & GraphDRP     & GratransDRP  & TransEDRP    \\ \hline
base                 & 0                  & 0                  & \XSolidBrush & 0.1935/.2073 & 0.0587/.4442 & 0.0637/.4402 & 0.0595/.4848 & 0.0624/.5060 \\ \hline
\multirow{2}{*}{(A)} & 70                 & 2                  & \Checkmark   & 0.0813/.3709 & 0.0549/.4946 & 0.0579/.4898 & 0.0561/.5103 & 0.0584/.5316 \\ 
                     & 70                 & 2                  & \XSolidBrush & 0.0544/.4852 & 0.0558/.4676 & 0.0577/.4788 & 0.0579/.4841 & 0.0590/.5143 \\ \hline
\multirow{2}{*}{(B)} & 15                 & 3                  & \Checkmark   & 0.0797/.3572 & 0.0567/.4760 & 0.0604/.4643 & 0.0561/.5034 & 0.0573/.5371 \\
                     & 10                 & 2                  & \Checkmark   & 0.0913/.3118 & 0.0572/.4699 & 0.0615/.4555 & 0.0575/.4950 & 0.0576/.5311 \\  \hline
\multirow{3}{*}{(C)} & 5                  & 1                  & \Checkmark   & 0.1203/.2585 & 0.0578/.4597 & 0.0630/.4456 & 0.0595/.4848 & 0.0588/.5177 \\
                     & 10                 & 1                  & \Checkmark   & 0.1134/.2773 & 0.0573/.4676 & 0.0621/.4524 & 0.0580/.4924 & 0.0580/.5273 \\ 
                     & 15                 & 1                  & \Checkmark   & 0.1115/.2927 & 0.0569/.4740 & 0.0614/.4581 & 0.0570/.4979 & 0.0591/.5323 \\ \hline
\end{tabular}%
}

\label{tab:2}
\end{table*}

\begin{figure*}[htpb]
    \centering
    \includegraphics[width=0.95\linewidth]{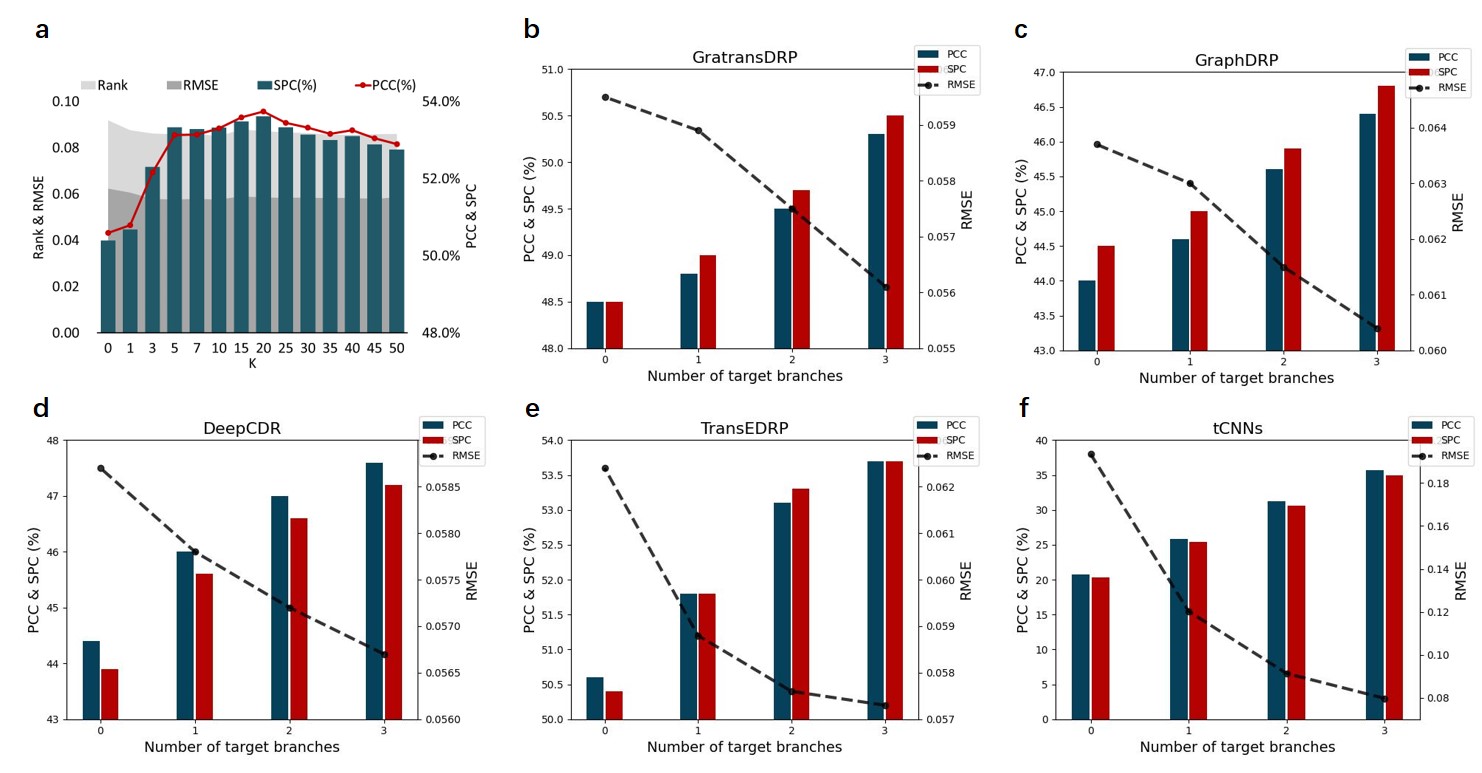}
    \caption{\textbf{Hyperparameter experiment.} \textbf{a}, The impact of the source domain consisting of the number of drug domains $K$ on the performance of the MSDA when the number of target domain adaptation branches is 2. \textbf{b-f}, The impact of the number of target domain adaptation branches $n$ on the performance of the MSDA when the number of drug domains corresponding to each target domain branch is fixed at 5. The experiments are conducted on the GDSCv2 public dataset, where \textbf{a} is the MSDA loaded on the TransEDRP method, and \textbf{b-f} are the MSDA loaded on five representative DRP methods. }
    \label{fig:Hyperparameter}
\end{figure*}

% integrates seamlessly with the public DRP methodology.
\subsection*{Ablation study}

In this study, we introduce the MSDA plug-in as a key component of our proposed solution, which aims to enhance the performance of drug response prediction in the context of zero-shot learning. This plug-in can be seamlessly integrated with publicly available DRP methods. The MSDA creates multiple target domain adaptation branches by replicating the fusion prediction branches from the pre-trained DRP model while preserving the parameters of the original model prediction branches. Finally, the MSDA combines the predictions from the target domain adaptation branches with those of the original branches, using summation and averaging, to generate the fine-tuning results specific to the target domain. Therefore, in the design of the structure of the MSDA, two crucial aspects necessitate verification through ablation experiments:

\begin{description}
\item[$ \mathrm{\mathbf{A1} }$  ] Whether the average adaptive fine-tuning of multiple target domain adaptation branches is better than that of just one single branch for $K$ drug domains as source domains filtered by the domain selector.

\item[$ \mathrm{\mathbf{A2} }$  ] Whether the results of the target domain adaptation branches $n$ need to be merged with the results of the original branch predictions.
\end{description}

For \textbf{A1}, to investigate the structural performance disparities between multi-branching and single-branching within the MSDA framework while maintaining an equal number of source drug domains, we fixed the number of source domains $K$ at 10 and 15, representing two and three times the original number of units (5), respectively. In the single-branching experiments, we set $K$ to 10 and 15, respectively. In the multi-branching experiments, the number of source domains in each branch is fixed at 5, while $n$ is set to 2 and 3, respectively. The ablation experiments of five DRP methods are conducted on the GDSCv2 public dataset. Table \ref{tab:2} (B) and (C) show the results of the experiments with an overall number of source domains of 15 and 10, respectively. These results indicate that a rational division of the $K$ domains into $\frac{K}{n}$ sub-source domains, followed by fine-tuning using $n$ target domain adaptation branches, is a more effective strategy. It also suggests that the MSDA may achieve better performance if the number of branches $n$ is increased.

For \textbf{A2},  we discuss the effectiveness of integrating the prediction branch of the original method with the target adaptation branch. The experiment is conducted using GDSCv2, where $K$ is set to 70 and $n$ is 2. We again select five representative DRP methods. As indicated in Table \ref{tab:2} (A), the performance is enhanced when the original prediction branches of each DRP model are integrated. This is because, during the inference stage of domain adaptation, the fine-tuned branch lacks access to real results from the target domain, making it challenging to gauge the extent of domain adaptation adjustments. Integrating the domain adaptation branch with the original branch stabilizes the prediction results for the target domain. Furthermore, it is observed that when the original method's effectiveness (e.g., tCNNs) is notably subpar, the mandatory fusion of the original branch and the domain adaptation branch diminishes the improvements brought about by the MSDA. Consequently, when employing MSDA as an adaptation adjunct during the testing phase, it is imperative to assess the performance of the incorporated method. 

\subsection*{Hyperparameter experiment}

The MSDA acts as a multi-branch multi-source domain adaptation test-time plug-in that can be integrated into general DRP methods, incorporating two critical hyperparameters:

\begin{description}
\item[$\mathrm{\mathbf{B1} }$ ] The number of drug domains $K$ selected by the domain selector while maintaining a fixed model structure.

\item[$ \mathrm{\mathbf{B2} }$  ] The number of target domain adaptation branches $n$ with the number of drug domains remaining fixed for each branch adapted to the target domain.
\end{description}

% \begin{itemize}
% \item  B1: The number of drug domains $K$ selected by the Domain Selector under the fixed model structure.

% \item  B2: The number of target domain adaptation branches $n$ under the condition that the number of drug domains corresponding to each branch adapted for the target domain is fixed.
% \end{itemize}

% These two hyperparameters are concerned with how to combine the performance improvement and inference speed of the MSDA. For this purpose, we conduct two experiments separately to explore these issues.

% For \textbf{B1}, we conduct a $K$ sensitivity analysis on GDSCv2 using the TransEDRP method with 14 parameters selected with varying spacing from 0 to 50. The trend of the four metrics with increasing $K$ in Fig. X shows that the effect of $K $on the MSDA method is more significant. When $K$ is equal to zero, that is, when we do not use the MSDA plug-in, the model performance is the lowest, when $K$ is from 5 to 20, the enhancement begins to remain stable, $K=20$ combines both inference speed and model performance, and when $K$ is greater than 20, the enhancement remains stable or there is a downward trend. This indicates that for the target drug, selecting too many drug domains as source domains at once will cause the MSDA method to decrease in the enhancement amplitude, and when most drugs and the target drug do not have a high degree of similarity, using their drug domains as source domains will confuse the valid information.

For \textbf{B1},  we perform a sensitivity analysis with $K$ values, considering 14 parameters selected at uneven intervals from 0 to 50, utilizing the TransEDRP method within the GDSCv2 dataset. The trend in Fig. \ref{fig:Hyperparameter}(a) illustrates the substantial impact of varying the value of $K$ on the four metrics. At $K=0$, representing the absence of the MSDA plug-in, the model performance is at its worst. Between $K=5$ and $K=20$, improvements stabilize. At $K=20$, a balance between inference speed and model performance is achieved. Beyond $K=20$, improvements either remain stable or exhibit a decreasing trend. This indicates that if the similarity between drugs in the source domains and those in the target domains is relatively low, confusion may be introduced into the pertinent information.

% It can be noticed from Table \ref{tab:2} which shows the prediction metrics of the five methods on GDSCv2 that the impact of increasing K on the performance is also relatively high when the number of branches is equal to 1.

% For \textbf{B2}, we set up 3 sets of experiments in sequence when 5 drug domains are used as source domains on each branch, the number of adaptation branches is $[1,2,3]$, and a blank control group is added. We performed prediction experiments on GDSCv2 for five existing publicly published methods. The experimental results in Fig. X show that the more the number of branches of the target domain adaptation, the more significant the improvement of the method performance of MSDA under the condition of a certain number of drug domains in each domain adaptation branch. We summarise that the strategy of MSDA should be to increase the number of target adaptation branches, $n$, as allowed by the number of model parameters, with each branch only touching $\frac{K}{n}$ drug domains from the source domain. This strategy allows MSDA to combine both performance and inference speed where arithmetic power permits, and to achieve stable and effective enhancement across different datasets and different methods.

For \textbf{B2}, we conduct three sets of experiments, each with a varying number of target domain adaptation branches ($n\in [1,2,3]$), while using five drug domains as the source domains on each branch. Additionally, a blank control group is included. The experimental results on GDSCv2 for five existing published methods in Fig. \ref{fig:Hyperparameter} (b-f) illustrate that as the number of target domain adaptation branches increases, the performance improvement of the models becomes more substantial for a specific number of drug domains that can be learned within each target domain adaptation branch. This strategy enables the MSDA to balance performance and inference speed within the constraints of available computational resources, resulting in consistent and effective improvements across various datasets and methods.

\section*{Conclusion}\label{sec3}

% 在这项工作中，我们提出了MSDA，一种测试时增强的即插即用的多源域适应的解决方案。MSDA是第一个针对DRP回归任务的多源域适应解决方案，将多个药物源域和目标域的药物反应信息的分布对齐并微调预训练模型以增强跨域泛化能力。我们在GDSCv2和NCI60数据集上实验，和其他最先进的DRP方法在相同训练集上预训练的结果相比，集成MSDA后的所有方法均有效地实现了跨域的泛化能力提高。此外，我们提出了一种新的指标HC，用于全面地直观地衡量模型在药物发现方面的性能，实验结果表明MSDA方案在HC上的效果显著。在这些实验的基础上，我们分析了xxx问题。这些结果表明MSDA有望作为在实际的药物发现领域中DRP任务上的一种重要的插件。

% 本文中，我们提出了一种可提高药物反应预测域外泛化性的解决方案。该解决方案中，我们提出了多分支的多源域适应的测试增强插件，叫做msda。MSDA可以无缝地加载在通用的drp方法上，利用现有药物反应数据库对无标签的类药化合物的预测结果实时修正。实验证明这种解决方案可以在临床前药物筛选阶段，普遍提高drp方法预测无标签化合物的精度。

% This paper presents a general solution to enhance the out-
% of-domain generalization for DRP models. In this solution, we propose a Multi-branch Multi-Source
% Domain Adaptation Test Enhancement plug-in, called MSDA. The MSDA can be seamlessly loaded on
% generic DRP methods to correct the prediction results of unlabelled novel compounds in real-
% time using existing drug response databases. Our experiments are conducted on the GDSCv2 and
% CellMiner datasets, and the results show that this solution can generally improve the generalization
% of DRP methods at the preclinical drug screening stage.

% 在现有的全局实验，消融实验和超参数实验的基础上。我们验证了域外泛化性增强方案中的MSDA插件具有即插即用，通用性强，泛化性提升显著，并且在算力条件允许的情况下具有更高潜能的特点。应对实际产业需求，MSDA可自适应地选择不同的参数以平衡性能和推理速度的需求。

In this paper, we present a zero-shot learning solution designed to cope with the DRP task in preclinical drug screening. We design the Multi-branch Multi-Source Domain Adaptation Test Enhancement Plug-in, called MSDA. The MSDA is seamlessly integrated into conventional DRP methods, enabling the learning of invariant features from previous response label information and thereby augmenting the model's real-time prediction capabilities.

We conduct a series of experiments on GDSCv2 and CellMiner datasets with the same dataset division standard and show that MSDA generally improves the performance of the DRP methods by 5-10\% during the preclinical drug screening phase. Among the five most representative and high-performing DRP methods, the incorporation of the MSDA significantly enhances their predictive performance in the zero-shot learning scenario. More specifically, the MSDA enhances the performance of each of the five methods (tCNNs, DeepCDR, GraphDRP, GratransDRP, and TransEDRP) by an average of 15.6\% (ranging from 9.3\% to 26.1\%) on GDSCv2 and by an average of 13.1\% (ranging from 1.5\% to 26.2\%) on CellMiner. The experimental results show that the MSDA plug-in can generally improve the prediction performance of the DRP methods at the preclinical drug screening stage for unlabeled compounds. Through our experiments, ablation studies, and hyperparameter analysis, we validate that the MSDA plug-in in our zero-shot learning solution is plug-and-play, highly versatile, able to enhance generalizability, and holds substantial promise for achieving superior performance while maintaining computational feasibility. To deal with the actual needs of the industry in real-world applications, the MSDA can adaptively select different parameters to balance the performance and inference speed requirements. 

% In this work, we present a general solution to discover more effective drug candidates from the unknown chemical space. In this solution, we propose the MSDA, a Multi-branch Multi-Source Domain Adaptation Test Enhancement plug-in, to enhance the zero-shot learning generalization for DRP models. The MSDA can integrate seamlessly with general DRP methods to correct the prediction results in real time. The MSDA is the first multi-source domain adaptation method for the DRP regression task that aligns the distributions of drug response information across multi-source drug domains and target domains, then fine-tunes the pre-training model to enhance the zero-shot learning generalization capability in the inference phase. 

% The MSDA improves the five methods (tCNNs, DeepCDR, GraphDRP, GratransDRP, and TransEDRP respectively) in the inference phase by the global average of 15.6\% (specifically 26.1\%, 15.8\%, 16.7\%, 9.3\%, 9.8\%) on GDSCv2 and 13.1\% (specifically 26.2\%, 15.6\%, 1.5\%, 10.2\%, 11.8\%) on CellMiner.

In summary, our zero-shot learning solution can effectively predict drug responses for compounds without label data, potentially expanding the scope of drug screening and improving the performance of the DRP methods in the preclinical drug screening phase. This contributes to streamlining drug discovery, resulting in substantial time and cost reductions. Our achievement—enhanced prediction accuracy in zero-shot learning—opens up possibilities for more efficient identification of prospective drug candidates, along with the potential for reduced experimental expenditure. It also underscores the opportunities for advancing research and innovation at the intersection of drug discovery and machine learning.

% 我们的解决方案可以对缺乏标记数据的化合物进行药物响应预测，潜在地扩大了药物筛选的范围，并且提高临床前药物筛选阶段DRP方法的性能5-10％，这有助于增强药物发现过程，实现显著的时间和成本节约。我们在提高预测准确性方面的成功可以更有效地识别有前景的药物候选，从而减少实验成本，突显了在药物发现和机器学习领域进一步研究和创新的潜力。
%
% In this work, we present the MSDA, a test-time-enhanced plug-and-play general solution for multi-source domain adaptation. The MSDA is the first multi-source domain adaptation solution for the DRP regression task that aligns the distributions of drug response information across multi-source drug domains and target domains, then fine-tunes the pre-training model to enhance the  zero-shot learning generalization capability. We experimented on the GDSCv2 and CellMiner, and all methods after integrating the MSDA effectively achieve improved generalization across domains compared to the results of other state-of-the-art DRP methods pre-trained on the same training set. In addition, we propose a new metric HC, to comprehensively and intuitively measure the performance of models in drug discovery, and the experimental results show that the MSDA scheme is effective on HC. On the basis of these experiments, we analyze the \textbf{xxx} problem. These results show that the MSDA is expected to serve as an important plug-in for the DRP task in the real drug discovery field.

\section*{Methods}\label{sec4}

\subsection*{Problem definition}
\label{sec4.1}

Preclinical drug screens seek compounds from a novel compound collection that have desired pharmacological effects when applied to specific cell lines. In the drug response prediction task in preclinical drug screening, the input data is the genomics of one compound molecule and one cell line, and the output is the half maximal inhibitory concentration ($IC50$ scores) of this molecule on one cell line. Since these compounds are absent from the drug response databases and are not visible during the model training phase, this constitutes a zero-shot learning problem. In the preclinical drug screens, we simulate real-world situations using the known drug response database. In a computer-assisted laboratory environment, we emulate real zero-shot conditions by employing actual response data from select drugs within the existing drug response database as our test dataset. The precise problem is delineated as follows.

A single drug domain $j$ is defined as a set of response data for a specific drug, $d_j$, across multiple cell lines, denoted as ${c_1^j, c_2^j, \ldots, c_n^j}$, comprising a total of $n$ samples. Each instance of response data, resulting from the interaction between a drug and a specific cell line, is considered one sample pertaining to that drug. The drug response data are divided into drug data domains based on drug types, denoted as $S=\left \{ (X_{sj}, Y_{sj}) \right \}_{j=1}^{N} \sim \left \{ p_{sj}(x,y) \right \}_{j=1}^{N}$; here, $\left \{ p_{sj}(x,y) \right \}_{j=1}^{N}$ are $N$ different source distributions. $X_{sj}=\left \{ (d_{j}^{s},c_{i}^{sj}) \right \}_{i=1}^{\left | {X_{sj}} \right |} $ represents samples from source domain $j$, while $Y_{sj} = \left \{ y_{i}^{sj} \right \}_{i=1}^{\left | {X_{sj}} \right |} $ is the corresponding $IC50$ scores of samples. $d_{j}^{s}$ represents one type of compound and $c_{i}^{sj}$ represent the cell lines from the source domain $j$. Each source domain consists of one type of compound, multiple types of cell lines, and corresponding $IC50$ scores. The compound $d_{j}^{s}$ is represented by the Simplified Molecular Input Line Entry System (SMILES) sequence or an undirected graph $G=(V,E)$, where $V$ denotes nodes and $E$ denotes undirected edges; the cell line $c_{i}^{sj}$ is represented by the genomics, including mutation and copy number aberration. The set of $M$ different novel compounds without the label can be denoted as $T=\left \{ (X_{tj}) \right \}_{j=1}^{M} \sim  \left \{ p_{tj}(x,y) \right \}_{j=1}^{M} $. As far as the drug discovery domain is concerned, the \textit{core requirement} of the DRP task is as follows: the model trained on $S$ achieves high accuracy on $T$ to verify that the model already has high generalization ability, which ultimately serves the practical application of drug discovery.

\subsection*{Shared drug and cell line feature extractors}
\label{sec4.2}
As elucidated in Section 2.1, the drug feature representation branch and the cell line gene feature representation branch share weights derived from pre-trained DRP models, which are not incorporated into gradient calculations. This paper does not provide a detailed exposition of the computational methods employed in these two data representation branches. Broadly, we categorize data feature extraction techniques into Convolutional Neural Networks (CNNs), Multi-Layer Perceptrons (MLPs), Graph Neural Networks (GNNs), Transformers, and Transformer-based GNNs.

The input types for the drug extractor are the SMILES sequence or molecular graph. We uniformly describe the process of branching feature representation of a drug $d^{\xi}_j$ as follows:
% $\mathbb{P}^t_i= \Phi _{drug}(d^{t}_i)$

\begin{equation}
\mathbb{P}^{\xi}_j= \Phi_{drug}(d^{\xi}_j)
\end{equation}
\noindent where, $\Phi_{drug}$  denotes the drug extractor with shared weights, while $\mathbb{P}^{\xi}_j$ is the representation of the drug $d^{\xi}_j$. $\xi = S/T $  refers to the drug domain $j$ that corresponds to this drug $d^{\xi}_j$ from the source or target domain dataset. Similarly, $c^{\xi j}_i$ is the input of the cell line extractor which can be described as follows:

\begin{equation}
\mathbb{Q}^{\xi j}_i= \Phi_{cell}(c^{\xi j}_i)
\end{equation}
\noindent where, $\Phi_{cell}$  denotes the cell line extractor with shared weights, while $\mathbb{Q}^{\xi j}_i$ is the representation of the cell line $c^{\xi j}_i$, $i\in [0,\left | X_{\xi j} \right | )$.

\subsection*{Multi-source domain selector}

Due to the vast number of drugs with known responses, utilizing the complete set of drug domains as source domains when implementing the MSDA for domain adaptation presents significant computational and time-related challenges. Therefore, a practical approach to mitigate computational resource constraints and enhance inference speed involves the selection of partially effective drug domains as source domains. The structure of the domain selector is illustrated in Fig. \ref{fig-2}(b). In the initial stage of the MSDA, we design the multi-source domain selector to select multiple drug domains that are similar to the target domain from the training dataset; these are defined as the source domains. 

The total number of source domains $S=\left \{ (X_{sj}, Y_{sj}) \right \}_{j=1}^{N}$; here, $d_{j}^{sj}$ represents the drug from the drug domain $j$ (as shown in Section \ref{sec4.1}). Each drug domain contains only one drug. Accordingly, the set of drugs from all source domains can be defined as $D^s=\left \{  d_{j}^{sj}  \right \}_{j=1}^{N}$, where $N$ is the total number of all the source domains. Similarly, the set of drugs in all target domains is denoted by $D^t=\left \{  d_{j}^{ti}  \right \}_{i=1}^{M}$, and $M$ is the number of target domains. In general, $M$ is a finite number in our experimental setting; in the real world, $M$ would be enormous. Next, for a target drug domain $D_{i}^t$, its multi-source domains $D^{s\to t}_i$ are selected as shown below:

\begin{equation}
D^{s\to t}_i= \mathrm{Top}_K(W_{ij}), \quad \mathrm{where}\ j\in [0,N )
\end{equation}
\noindent where $\mathrm{Top}_K$  denotes the operation of sorting from smallest to largest order and fetching the first $K$ elements. The Wasserstein distance $w_{ij}$ is a distance function defined between probability distributions on a given metric space, and its 1-order form is formulated as follows:

\begin{equation}
W_{ij}= W(\mathbb{P}^s_j ,\mathbb{P}^t_i ) = \underset{\gamma \in \Gamma  }{\mathrm{inf} } \mathbb{E}_{(x,y)\sim \gamma } \left [ \left \| x-y \right \|  \right ] 
\end{equation}
\noindent  where $\Gamma  = \Pi (\mathbb{P}^s_j ,\mathbb{P}^t_i) $ denotes the set of all joint distributions $\gamma(x,y)$ whose marginals are respectively $\mathbb{P}^s_j= \Phi _{drug}(d^{s}_j)$ and $\mathbb{P}^t_i= \Phi _{drug}(d^{t}_i)$; here, $\Phi _{drug}(\cdot)$ is the shared drug feature extractor. The distributions of various drug features exhibit significant dissimilarity, often with minimal overlap. The advantage of using the Wasserstein distance over the Kullback-Leibler (KL) and Jensen-Shannon (JS) divergences lies in the former's ability to capture the proximity between two distributions, even when there is no overlap between them. This property enables the measurement of distance between the source and target domains, even in scenarios involving non-overlapping drug distributions.

% Distributions of different drug features are very different, and to a large extent there may be only minimal overlap. The superiority of the Wasserstein distance over the Kullback-Leibler (KL) divergence and Jensen-Shannon (JS) divergence is that even if the two distributions do not overlap, the Wasserstein distance still reflects their proximity. This makes it feasible to measure the distance between the source and target domains of even partial drugs.

% $\mathbb{P}^s_j$($\mathbb{P}^t_i$) is the representation matrix of $d^{s}_j$($d^{t}_i$) computed by $Drug$ is the shared drug feature extractor.

\subsection*{Multi-branch drug domain adaptation}

The multi-branch drug domain adaptation module includes three functions:  the fusion of drugs and cell line features, the adaptation from the multi-source domains to the target domain, and the predictions of drug responses in the target domain. 

The inputs of the drug and cell line feature fusion branches are the outputs of the shared drug feature extractor $\Phi_{drug}$ and the shared cell line feature extractor $\Phi_{cell}$, respectively (refer to Section \ref{sec4.2} for details). Specifically, this module has a fusion branch for multi-source domains and several fusion branches for the target domain. The fusion branch can be expressed as follows:

\begin{equation}
    Y_{s\to t}^{pred} = \Phi_{fusion}( \mathbb{F}^{s\to t} )
\end{equation}
\noindent where, $\mathbb{F}^{s\to t}=\left [ \mathbb{P}^{s\to t}, \mathbb{Q}^{s\to t} \right ]$, and $\Phi_{fusion}$ denotes the fusion branch that can be replaced by different methods. The fusion branch of the multi-source domains uses pre-trained model parameters and is not involved in the gradient computation.

Each target domain fusion branch is constrained by three conditions, which are regression loss and ranking loss in multi-source domains $D^{s\to t}$ and feature-invariant consistency based on the MMD distance between $D^{s\to t}_i$ and $D^{t}_i$. The objects of the feature-invariant constraint are $FC(\mathbb{F}^{s\to t}_i)$ and $FC(\mathbb{F}^{t}_i)$, which are activated by the fully connected layer $FC(\cdot)$ once. The formulas for these constraints are as follows:

\begin{equation}
\mathcal{L}_{reg} =  {\sum_{i=0 }^{\left | D_i^{s\to t} \right | } \mathrm{MSELoss}(Y_{i}^{pred},Y_{i})} 
\end{equation}
\noindent where, $MSELoss(\cdot)$ denotes the L2 norm loss. The label value $y_i$ (IC50 score) is subtracted from the model output (estimate) $f\left ( x_i \right )$, after which the square is calculated to obtain the L2 norm loss, which is expressed as follows:

\begin{equation}
\mathrm{MSELoss}\left ( x,y \right )  = \frac{1}{n}  {\textstyle \sum_{i=1}^{n}} \left ( y_i-f\left ( x_i \right )  \right ) ^2
\end{equation}

The ranking loss $ \mathcal{L}_{rank} $ is computed using MarginRankingLoss (MRLoss). For data $(x_1, x_2, r)$ containing $N$ samples, $x_1$ and $x_2$ are the two inputs given to be ranked, and $r$ represents the true ranking labels. The ranking loss $ \mathcal{L}_{rank} $ is computed as shown below:

\begin{gather}
\mathcal{L}_{rank} =  {\sum_{i=0 }^{\left | D_i^{s\to t} \right | } \mathrm{MRLoss}(Y_{i}^{pred},Y_{i})}
\\
\mathrm{MRLoss} = \max(0, -r(x_1 - x_2) + \mathrm{margin})
\end{gather}
\noindent where $ \mathrm{margin}$ denotes the margin value. If this value is larger, it means that the expectation $x_1$ is further away from $x_2$ (i.e. the margin is larger). The MMD (Max mean discrepancy) is one of the most widely used loss functions in transfer learning, especially in domain adaptation, and is mainly used to measure the distance between two different but related distributions. The distance between two distributions is defined as follows:

\begin{equation}
    \mathrm{MMD } (\mathrm{X} ,\mathrm{Y} ) = \left \| \frac{\sum_{n}^{i=1} \phi\left ( x_i \right )}{n} -\frac{\sum_{m}^{j=1} \phi\left ( y_j \right )}{m} \right \|^2_H
\end{equation}
\noindent where, $\phi(\cdot)$ denotes a mapping from the original space to Hilbert space $H$. Hilbert space is the extension of Euclidean space that is no longer restricted to the finite-dimensional situation. The dimensions of the drug and cell line features encoded by the different methods are different. The fact that the calculation of the distance between two distributions is not limited by the dimension of the sampled features is one of the keys to the functioning of the MSDA as a general plug-in. The distribution distances of invariant features from between multi-source domains $D^{s\to t}_i$ and the target domain $D_i^t$ are constrained by MMD distances as follows:

\begin{equation}
    \mathcal{L}_{mmd} =   {\sum_{i=0 }^{\left | D_i^{s\to t} \right | } \mathrm{MMD } \left (\varphi_{\mathrm{align}}  \left ( \mathbb{F}^{s\to t}_i , \mathbb{F}^{t}_i \right )  \right )    }
\end{equation}
\noindent where $\varphi_{\mathrm{align}}(\cdot)$ denotes the operation in which the feature vectors of two distributions are first fused through a fully connected layer, after which a union set is taken according to the cell line types, and eventually, the features are aligned according to their cell line types. 

Finally, the overall loss of each target domain adaptation branch is obtained by weighted summation, as follows:

\begin{equation}
    \mathcal{L} = \mathcal{L}_{reg} + \alpha \mathcal{L}_{rank} +  \beta   \mathcal{L}_{mmd} 
\end{equation}
\noindent where, $\alpha \in [0,1]$ and $\beta \in [0,1]$ are the weights of the loss function. The computation of the loss of each target domain adaptation branch, the back-propagation, and the updating of the gradient are independent and serial.

\subsection*{Experimental settings}

\subsubsection{Baseline models}

Several existing representative methods are selected as the baseline models, including 1DCNN-based, graph-based, and transformer-based methods. The selected graph-based methods are GraphDRP \cite{mainDRP3} and DeepCDR \cite{DeepCDR}; the selected 1DCNN-based method is tCNNs \cite{tcnns}; finally, the selected transformer-based methods are  GraTransDRP \cite{GraTransDRP} and TransEDRP \cite{TransEDRP}, which also encode drug molecules as graphs for the initial representation of molecules.

\subsubsection{Datasets}

We evaluate the MSDA with five SOTA baselines
on two public DRP datasets: GDSCv2 \cite{GDSC} and CellMiner \cite{NCI60}. The GDSCv2 dataset is a web-accessible database, which is an academic research program to identify molecular features of cancers that predict response to anti-cancer drugs.  GDSCv2 contains 1000 human cancer cell lines, is screened with hundreds of compounds, and is the most commonly used dataset for DRP tasks. CellMiner is a database and query tool designed for the cancer research community to facilitate the integration and study of molecular and pharmacological data for the NCI-60 \cite{NCI-60_1, NCI-60_2} cancerous cell lines. 

% The NCI-60 is a panel of 60 diverse human cancer cell lines used by the Developmental Therapeutics Program of the U.S. National Cancer Institute to screen over 100,000 chemical compounds and natural products.

\section*{Data availability}
The original GDSCv2 and CellMiner data are publicly available
datasets.  GDSCv2 is downloaded from the website (\url{https://www.cancerrxgene.org}). CellMiner is downloaded from the website (\url{https://discover.nci.nih.gov/cellminer/home.do}).

\section*{Code availability}
We load the MSDA on various SOTA DRP methods. The source code is available at \url{https://github.com/DrugD/MSDA}.

\bibliography{refs}
% \bibliographystyle{IEEEtran}
% \bibliography{refs}

\section*{Acknowledgements}
Wenbin Hu is supported by the Natural Science Foundation of China (Nos. 61976162, 82174230), Artificial Intelligence Innovation Project of Wuhan Science and Technology Bureau (No.2022010702040070), and Joint Fund for Translational Medicine and Interdisciplinary Research of Zhongnan Hospital of Wuhan University (No. ZNJC202016).

\section*{Supplementary Material}
Supplementary Information is available for this paper.

\end{document}